# Ceramics with metallic lustre decoration.
## A detailed knowledge of Islamic productions from 9th century until Renaissance

Delhia Chabanne[1,3], Marc Aucouturier[1*], Anne Bouquillon[1], Evelyne Darque-Ceretti[2]
Sophie Makariou[4], Xavier Dectot[5], Antoinette Faÿ-Hallé[6], Delphine Miroudot[4]

[1] C2RMF (UMR CNRS 171), Palais du Louvre, 14 quai François Mitterrand, 75001 Paris ;
marc.aucouturier@culture.gouv.fr, anne.bouquillon@culture.gouv.fr ; tel. +33-(0)1 4020 5749
[2] MINES-ParisTech, CEMEF (UMR CNRS 7635), BP 207, 06904 Sophia-Antipolis cedex, France ;
evelyne.darque-ceretti@mines-paristech.fr ; tel. +33-(0)4 9395 7454
[3] delhia.chabanne@gmail.com
[4] Musée du Louvre, DAI, 75001 Paris, France ; sophie.makariou@culture.gouv.fr,
delphine.miroudot@louvre.fr ; tel. +33-(0)1 4020 5069
[5] Musée national du Moyen-âge, 6 pl. Paul Painlevé 75005 Paris ; xavier.dectot@culture.gouv.fr ;
tel. +33.(0)1 53 73 78 00
[6] Previously at Musée national de la Céramique, 92310 SEVRES, France ;
antoinette.halle@wanadoo.fr.
* corresponding author

**FOREWORD**: This article gathers the results of experimental measurements completed during years 2005-2006 and interpreted during years 2007-2008.
Because of various edition problems, they could not be published earlier than now (end of 2010 - beginning of 2011).

## 1. Introduction

It is a common statement that productions of ceramic art and craft industry are elected testimonies of the past civilisation. Indeed, starting from a material as ordinary and "primary" as earth, any enrichment and creation, any utilisation testifies from the technological progress and from the material and spiritual needs of a period, a period which can be identified because clay keeps the memory of places and times.

On the 9th century, during the most brilliant period of Islamic civilisation in Mesopotamia, under the Abbasid caliphate, appears an outstanding technique of ceramic decoration: *lustre*, a precursory nanotechnology, a true alchemy which is able to transform simple earth into infinitely precious objects, giving them magnificent shines including the appearance of gold [**1-3**]. That kind of decoration is related to a very sophisticated process which creates on the surface of a glazed ceramic a layer of vitreous matter with sub-micron thickness containing metallic particles (copper and silver) with a nanometric diameter [**3**]. It confers to the surface a particular coloured aspect, often metallic in specular reflection. The fabrication and use of lustred ceramics were propagated across the Islamic world as far as Spain, to lead to the creation of the Italian majolica at the Renaissance period [**1,4**].

Many researches were dedicated during the last decades to the circumstances of propagation of that technique during centuries [**1**], to the iconographical, typological and analytical description of the various production observed in the Islamic world [**2,5**] and to the classification and reproduction attempts of the different fabrication recipes listed in known documents of the Moslem tradition or in pottery treatises [**1-3,6**]. Others were devoted to the characterisation of a limited number of objects coming from excavations or kept in collections [**5,7**], to the interpretation of the very particular optical properties which bring them their peculiar aspect [**8,9,10**].

Our aim is not to add supplementary data to the now well established knowledge (see [**4,5,7**] and references therein included) concerning the fabrication process principle or the general structure of the surface layers of ceramics with a metallic lustre decoration. It is more to attempt a study, based on structural criteria as precise as possible, of by which ways that particular know-how conquered first the Islamic world and later south Europe starting from the initial radiating source and which transformations this know-how may have undergone in its route. This could be done thanks to the kind opening of their collections by the curators and staffs of several national French museums: the Islamic Art Department (DAI) of the Louvre museum; the *Musée national du Moyen-Age*; the *Musée national de Céramique de Sèvres*. The concerned period and geographical range starts from the 9$^{th}$ century Mesopotamian production and end with the Hispano-Moresque objects fabricated in the Spanish east coast workshops during the 14$^{th}$ to 17$^{th}$ centuries (plus some plates produced near Valencia in Spain during 18$^{th}$ century), including the production of a number of Islamic centres in Egypt, Syria, Iran, North Africa and Spain. More that 110 objects (entire pieces or shards) coming from those museum collections were characterised. This work does not take into account the optical properties of the lustre layers and their modelling which are conducted by researchers of a friend laboratory, the "Institut des NanoSciences de Paris", with their own competence [**8,9**].

The present article shows that it is possible to establish criteria for differentiating the productions, thanks to the use, in addition to characterisation methods commonly used in the numerous published works, as scanning electron microscopy, X-ray diffraction, etc. (see [**4,5,7**] and references therein included), of original powerful non-destructive methods for the determination of the structure and composition of surface layers.

**2. Presentation and short history of the lustred ceramic. Studied corpus**

The process steps to obtain a glazed ceramic with lustre decoration are as follows [**1-3**]: after the usual two firing sequences (high temperature firing of the ceramic body, then application and firing at intermediate temperature of the vitreous glaze, coloured or not), an additional treatment of annealing at moderate temperature (ca 500-600° C) is performed in presence of a clayey mixture containing, among others, metallic salts and organic compounds (*fig. 1*). This latter treatment is operated in a partially reducing atmosphere and leads to the formation of metal nanoparticles which remain imbedded in a thin surface layer of vitreous glaze.

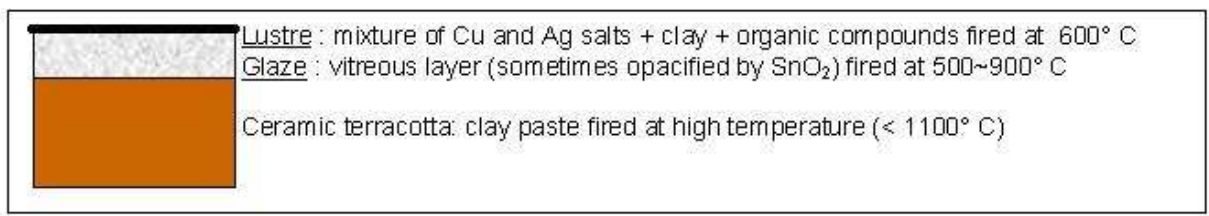

*Figure 1*: fabrication principle and structure of a glazed ceramic with lustre decoration

The technique appears initiated for glazed ceramics at the 9$^{th}$ century in the Mesopotamian area [**2**] under the Abbasid dynasty (750-1055), probably in Samara, Susa, Baghdad and Basra. That type of ceramic was fabricated and used under the great Islamic dynasties [**2**]. First, in Egypt, more precisely in Fustat, where an important production appeared under the Fatimids (969-1171), with perhaps some preceding examples in that region (said as pre-Fatimids). Then the technique diffuses in all Orient: in Syria (mainly Raqqa) under the Ayyubids (1171-1250) and during the Mamluk period (1250-1517); in Iran (notably in Kashan or Rayy), first under the Seldjukids (1038-1194) and Ilkhanids (1256-1353), then under the Timurids (1370-1506) and Safavids (1501-1732) with some examples (a renewal?) under the Gadjars (1779-1924). Simultaneously, the technique appears in Occident, in southernSpain as soon as the taifa merged after the dissolution of the Spanish Umayyad caliphate and long after, under the Nasrid dynasty (1237-1492) ; its apogee, in the 14$^{th}$ – 15$^{th}$ centuries gave rise to the Hispno-Moresque ceramic that was elaborated during five centuries later in the Valencia region until the 18$^{th}$ century. The technique finds a new application in Renaissance Italy (15th and 16th century) where Deruta and Gubbio become the most famous production centres of lustred glazed majolica [**4**].

The first Abbasid examples show a real mastering of the technique with in particular the fabrication of polychrome decorations (*fig 2A and 2B*). Later, polychromy was progressively lost (*fig. 2D*) but coloured glazes (blue, green, aubergine, etc.) or over-applied colour may be associated to the lustre (*fig. 2C and 2E*). The latest examples (18th century), in Safavid Iran or in eastern Spain exhibit decoration with a coppered aspect (*fig. 2F*).

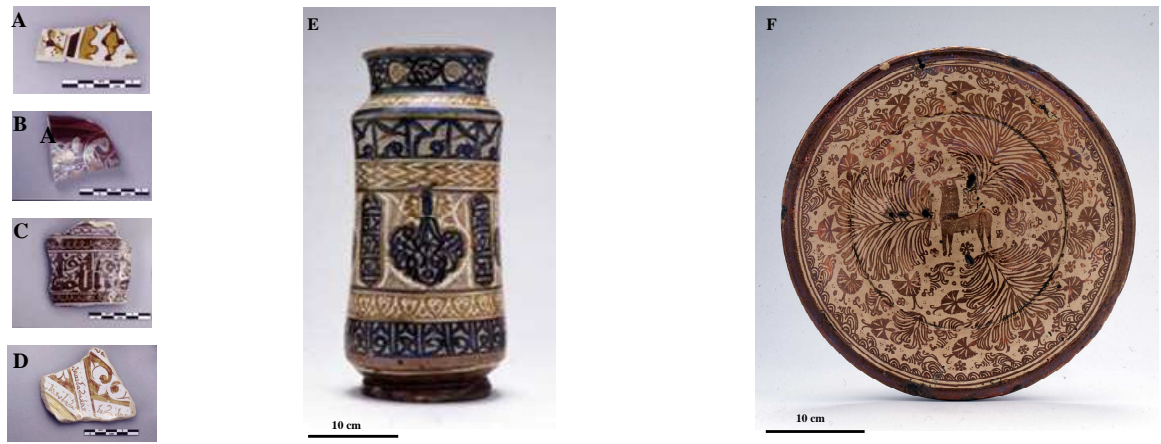

*Figure 2*: *examples of the studied objects (whole pieces or fragments): Abbasid period (A & B, © C2RMF, D. Bagault), Fatimid period (C & D, © C2RMF, D. Bagault) and Hispano-Moresque period (E & F, © Musée du Moyen-Âge, Cluny)*

A collaboration was established with three French national museums (Louvre museum, Musée national du Moyen Age and Musée national de Céramique de Sèvres) and a private collector. They provided the analysed objects (plates, vases, fragments, etc.) issued from the previously cited productions (**table 1 and appendix 1**). Particular care was given to the choice of the specimens as their decoration should be characteristic of a production (and neither coming from isolated attempts nor from failures; this is especially true for shards). Moreover, the choices contain the available different technological options of a given period (nature of the terracotta and of the glaze). To that list were added (also included of **table 1**) a set of objects studied in the frame of a collaboration with a researcher of Seville university [**11**] sponsored by the European program Eu-ARTECH, and, for comparison 2 pieces produced by a modern Spanish artisan [**12**].

Altogether 124 objects have been analysed during that program (**fig. 3**). The results may also be compared with results obtained on Italian majolica in the frame of a former study [**4**] done in the present laboratory in collaboration with the Italian CNR, the Gubbio museum (Italy), the Art Objects department of the Louvre museum, the *Musée national de la Renaissance* (Ecouen, France) and the *Musée national de Céramique de Sèvres*.

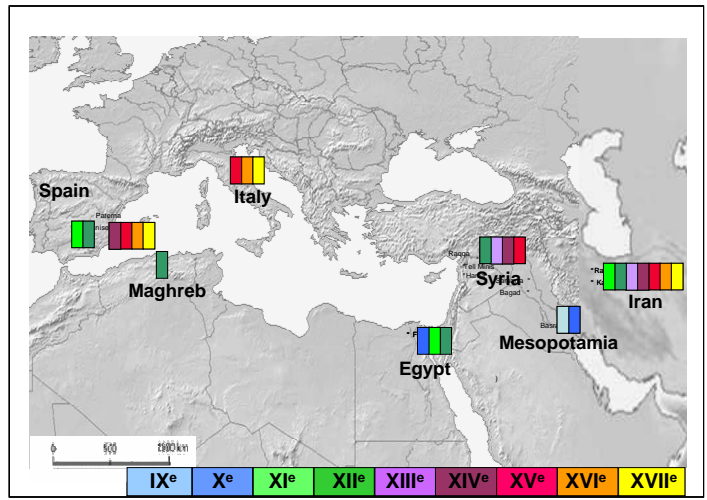

*Figure 3*: origin and dating of the studied objects

*Table 1: summing up of the glazed ceramics with lustre decoration studied in 2005-2006 (entire objects or fragments).*

| Period | Production site | Number of specimens | Conservation place |
|---|---|---|---|
| Abbasid 9$^{th}$ 10$^{th}$ cent. | Susa, Mesopotamia | 23 | DAI, Louvre museum |
| 9$^{th}$ – 11$^{th}$ cent.* Abbasid? | Mesopotamia? (*found in Fustat*) | 8*,*** | Private collection |
| Pre-Fatimid* | Fustat, Egypt | 4* | DAI, Louvre museum |
| Fatimid (989-1171) | Fustat, Egypt | 6(+4*) | Musée national de Céramique, |
|  |  | 5(+1*) | DAI, Louvre museum |
|  |  | 7(+1*) | DAI, Louvre museum |
| Ayyubid (1171-1250) | Syria | 2 | Musée national de Céramique, |
| Tell Minis 12$^{th}$ cent. | Syria | 1 | DAI, Louvre museum |
| Pre-Mongol (1038-1194) | Iran | 12 | DAI, Louvre museum |
| Mongol (1256-1353) | Iran | 10 | DAI, Louvre museum |
| Timurid (1370-1506) | Iran | 4 | DAI, Louvre museum |
| Safavid (1501-1732) | Iran | 6 | DAI, Louvre museum |
| Mamluk (1250-1510) | Syria | 2 | DAI, Louvre museum |
| End of 11$^{th}$ cent. | La Qala de Banu Hammad, Algeria | 3 | DAI, Louvre, UCAD deposit |
| Islamic Spain, 12$^{th}$ – 14$^{th}$ centuries | Andalusia, Spain | 3* | DAI, Louvre museum |
| Hispano-Moresque 15$^{th}$-18$^{th}$ centuries | Valencia and Seville, Spain | 10 | Musée du Moyen-Âge, Cluny |
|  |  | 3 | Musée National de Céramique, |
|  |  | 7** | Triana workshop, Seville |
| Modern | Granada, Spain | 2 |  |

* uncertain attribution
** kindly supplied by A. Polvorinos del Rio, Seville university [**11**]
*** kindly supplied by A. Kaczmarczyk [**15**]

## 3. Experimental approach and procedure

A detailed knowledge of the surface layers responsible for the lustre aspect needs a multiple method approach as less invasive as possible for those cultural heritage artefacts [**13**]. Observation at different scales by optical microscopy, conventional and high resolution scanning electron microscoy (SEM and HR-SEM), atomic force microscopy (AFM) and transmission electron microscopy (TEM) on cross section reveals their fine microstructure (**fig. 4**) [**14,15**].

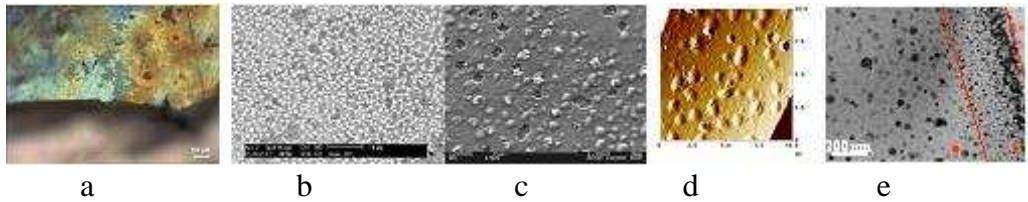

a       b       c       d       e

*Figure 4: observation of a lustred ceramic at different scales: (a) optical microscope; (b) conventional SEM; (c) HR-SEM; (d) AFM; (e) TEM*

Tiny metal particles are visible, with an average diameter of 10 to a few tens nanometres, sometimes agglomerated as clusters with a larger size. In the present case, the HR-SEM observation show that the particles are in fact lying under a surface film with a vitreous-like aspect sometimes broken, confirming observations reported in the literature [3,7,14].

The size and the nature of the particles may be measured by grazing incidence X-ray diffraction. On the diagrams of **figure 5** a broadening of the metallic silver diffraction peaks is observed. Through the application of the Sherrer formula [16], a measure of that broadening, corrected by the instrumental broadening obtained on a standard recrystallised silver specimen, leads to the calculation of the size of the smallest silver particles. In the shown example, the estimated size is 12 nm.

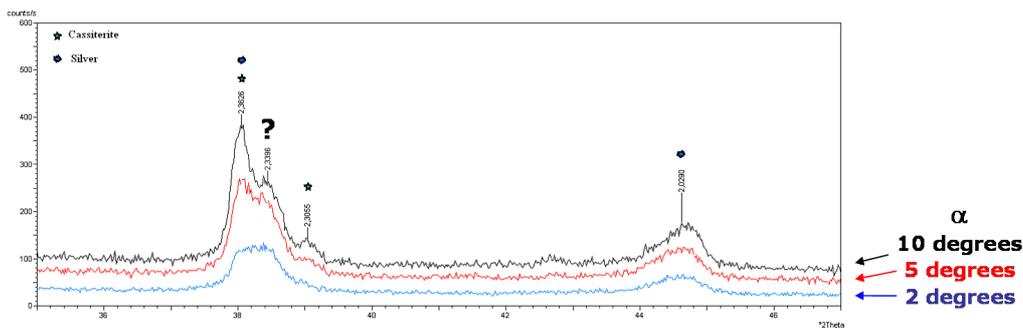

*Figure 5: grazing X-ray diffraction on a modern lustre surface ($\alpha$ = incidence angle)*

The aim of the present study was to investigate the eventual technological evolution of the lustre fabrication, through a comparison of the composition and structure of the ceramics and lustre layers coming from the different production centres.

Three main criteria have been held for that comparison:
- The nature and structure of the ceramic bodies;
- The composition and microstructure of the glazes;
- The thickness and composition of the different surface layers which constitute the actual lustre.

In order to quantify those criteria, several laboratory methods where used, with a preference given to non-destructive analyses methods. Practically all objects where analysed through ion beam analyses techniques on the 4 MV peletron particle accelerator AGLAE (NEC inc.) with an extracted beam available in C2RMF (Centre de Recherche et de Restauration des Musées de France) [17]: PIXE spectrometry under a 3 MeV energy proton beam associated with the quantification code GUPIX [18] for characterisation of the elemental chemical composition (major, minor and trace elements); and RBS under a 3 MeV alpha particle beam, which allows the quantitative determination of the in-depth composition profiles over a few micrometers from the surface; the particle beam (diameter lower than 50 μm) is extracted to free atmosphere in an helium flux by crossing a $Si_3N_4$ window of 100 nm thickness.

The RBS spectra are interpreted by using the SIMNRA calculation code [19]: a virtual specimen is built, constituted of a discrete number of superposed layers whose compositions and thicknesses are adjusted until the simulated spectrum fits the experimental spectrum. **Figure 6** shows an example of the followed process during the simulation and comparison of the result with observation in the transmission electron microscope of the cross section obtained on a similar archaeological specimen [20]: the spectrum of the bare glaze, considered as homogeneous in depth, is simulated thanks to the knowledge of its composition through PIXE analysis; then a glaze layer containing adequate contents of copper and/or silver is added to the top surface of the virtual specimen to try to simulate the experimental spectrum (**fig. 6a**); one observes that the energies of the peaks corresponding to both Cu and Ag may not fit with these observed; it is then necessary to add to the virtual specimen a layer containing neither of the metals, whose thickness is adjusted to provoke the adequate energy shift (**fig. 6b**); to achieve the simulation, one has to insert over or under the principal metal-containing layer one or several intermediate layers with decreasing Cu and/or Ag contents in order to simulate concentrations gradients before reaching the glaze (**fig. 6c**). The result appears in the scheme of **figure 6d**. Comparison with TEM observation (**fig. 6e**) shows that the RBS simulation result is in agreement with the actual structure, and this validates the methodological process.

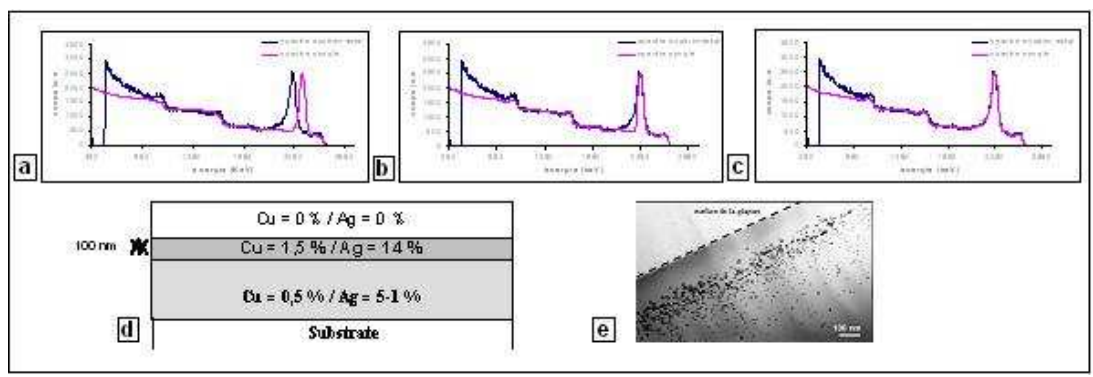

*Figure 6*: *experimental (curves with noise) and simulated RBS spectra on a specimen of lustred ceramic; (a) simulation of the glaze covered with a layer containing Cu and Ag; (b) addition of a layer containing neither copper nor silver; (c) insertion of an intermediary layer with a Cu and/or Ag gradient; (d) simplified result of the simulation; (e) TEM image of a cross section of a similar specimen (© CEMES-CRPAA, P. Sciau).*

It must be kept in mind that the representation of the subsurface distribution as a succession of discrete layers, a consequence of the simulation procedure of the SIMNRA code, is schematic. In fact, as shown in TEM images of **figures 4 and 6**, the distribution of the metallic nanoparticles as well as the concentrations of other elements vary continuously as a function of the depth.

In complement, observations and quantitative microanalyses have been done by optical microscopy and microanalytical scanning electron microscopy (SEM-EDS, Philips controled vacuum scanning electron microscope) on microsamples when sampling was possible (stratigraphic cross-sections imbedded in resin et polished down to ¼ μm diamond paste), in order to precise the nature, the composition and the microstructure of the terracotta and the glazes. Punctually, X-ray diffraction (Brücker D8 diffractometer) brought necessary structural information. Some objects (in particular the modern ones) were studied by high resolution scanning electron microscopy (HR-SEM, with a field emission gun, Leo 1450VP - SEM240, Ecole des Mines de Paris), by AFM and by grazing incidence X-ray diffraction, in order to precise the structure of the extreme surface.

## 4. Results and discussion

### 4.1. The pastes (terracotta)

**Table 2** gathers the measured average compositions of the terra cotta, obtained by PIXE analysis. The results do not concern all objects: it was not always possible to measure the terra cotta composition, especially for museum entire objects where no sampling was possible, and where the ion beam could not reach the clean paste body.

Two kinds of substrates are evidenced: in majority marly clays, beige to orange, and on the other hand highly siliceous pastes, white, near from archaeological faïence.

The oldest objects (Abbasid period) found in Susa and Samara are made from a homogeneous body, a marly (20 % CaO) ferrous (7 % $Fe_2O_3$) clay (**table 2**). This kind of paste is typical of productions from this region as similar chemical compositions are already found in the $4^{th}$ millenary BC for Susa I decorated potteries [**21**]. It is related to the local earth, common and approved. The paste is clear enough despite a relatively high iron content, because that element is incorporated during firing in pyroxene-type phases, avoiding the red colouration due to "free" iron. On all studied Abbasid potteries, no noticeable variation is observed, except varying contents in chlorides and sulphides attributed to burying salts or restoration products. That kind of marly clay is also found for pre-fatimid and part of the fatimid (between $9^{th}$ and $12^{th}$-$13^{th}$ centuries) potteries, confirming previous studies [**22**]. It is present later in Iran of Timurid era and is exclusively used for the Spanish productions.

On the other hand, the substrate is always siliceous in Syria during Ayyubid and Mamluk dynasties and in Iran for the pre-Mongol and Mongol productions.

Both kinds of paste are used during the Fatimid period and in Iran up to the $14^{th}$ century. The siliceous paste is constituted mainly of quartz grains embedded in an often alkaline matrix phase, following a tradition of antic Egyptian productions. It is however somewhat different by its higher alumina content (7 %), a proof of introduction in the mixture of small quantities of clay added to the alkaline flux.

As for the Safavids from the $17^{th}$ century onwards, not mentioned in **table 2**, it seems that both materials coexisted but this last statement needs to be confirmed, because the number of results is not large enough to be considered as valid.

**4.2. The glazes (table 3)**

Glazes are distributed into two classes: alkaline glazes always transparent and lead-bearing glazes generally opaque. Among the latter one may distinguish between those containing low tin amount (less than 10 wt % $SnO_2$) and those heavily opacified (about 20 wt % $SnO_2$).

The transparent glazes are mostly alkaline. With a somewhat high potassium content ($\approx$ wt % $K_2O$), colourless or greyish, they are typical of the first Abbassid productions. Under the Ayyubids and Safavids, glazes contain more sodium ($\approx$ 4-5 wt % $Na_2O$) and are sometimes coloured in blue or light green by cobalt and/or copper oxides (not mentioned in **table 3**). In the workshops of the Spanish Levant after the $17^{th}$ century colourless glazes with lead but without tin are applied, leaving the orange colour of the terra cotta visible.

The opaque glazes contain always lead, with PbO contents varying from 5 wt % under late Abbassids to 40 wt % under the Fatimids and then in Spain. Opacification is obtained by a more or less high amount of tin, between 4 and 20 wt % $SnO_2$. The whitest pre-Fatimid and early Fatimid ($9^{th}$-$10^{th}$ centuries) ones contain the highest tin content.

Some background glazes are present as well as highlights coloured in violet, blue or green by respectively manganese (0.3 to 0.7 wt % MnO), cobalt (up to 0.3 wt % CoO) and copper (up to 1 wt % CuO).

The material choice for the terra cotta and the glaze is a consequence of ceramic traditions often thousand years old, generally linked to the geological context, sometimes also to aesthetical or technical constraints. One must keep in mind that alkaline glazes do not fit easily clayey pastes [**23**]. For instance in Syria, the agreement between the paste and the alkaline glaze is obtained with an always siliceous substrate.

### 4.3. The lustre composition and structure

As stated in section 3, the non-destructive analysis procedure followed in this study gives access to the distribution of elements in the first layers of the objects under the actual surface. The SIMNRA simulation used to interpret the RBS spectra (**fig. 6**) describes these first layers in a simplified diagram of stacked discrete layers, although reality is the existence of continuous variations of elements. Such a simplified description gives nevertheless a good evaluation of the most specific physico-chemical features of a given lustre. **Figure 7** shows three examples obtained on specimens issued from different productions; the thicknesses of the different layers and their copper and silver nanoparticle contents as quantified the RBS spectra simulation are indeed in good agreement with cross section TEM observations.

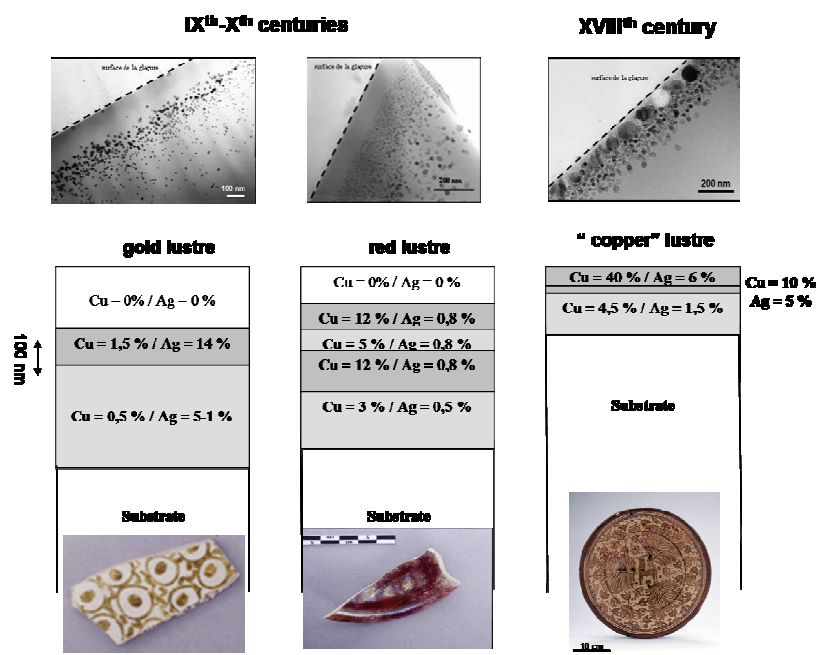

*Figure 7*: three lustre structures observed in TEM and corresponding results by RBS analysis simulation. TEM micrographs kindly provided by P. Sciau, CEMES-CNRS, Toulouse. Concentrations are in at. %.

Following that evaluation mode, lustre can be schematically characterised by a succession of more or less thin layers, not always present in their totality:
- An extreme surface layer of glaze containing neither copper nor silver, not always present;
- A set of layers describing a concentration gradient of copper and/or silver;
- A "main layer" containing the maximum copper and/or silver amount;
- Again a set of layers with a concentration gradient of copper and/or silver, extending until the metal-free glaze substrate.

The presence of absence of those layers, their thicknesses and their copper and silver content were taken as criteria for the comparison of the various productions and the detection of possible evolutions between them. Another criterion has also been considered: the *volume fraction* of copper and silver in each layer, more useful for interpreting and modelling the optical properties [**8,9**]. This fraction has been calculated by considering that both elements are in metallic form. It is known from the literature [**24**] that this assumption is not entirely satisfied for copper, which can be present as both metallic and oxide state in the lustre; this means that the copper volume fractions may be here overestimated.

Instead of giving a full and unreadable table of all the data, the following figures (**fig. 8 to 11**) tend to summarise the most characteristic features which may bring evidence of the evolutions in the manufacturing techniques. A detailed analysis of all the results lead to choose the following criteria found to be useful to attempt a comparison amongst the productions:
- The "surface layer thickness" describes the presence or absence of an extreme surface layer containing neither copper nor silver and its thickness when it exists; this parameter is important to compare the present results to observations usually reported in the literature (see for instance [4, 5, 7, 14]).
- The "total thickness" is the sum of thicknesses of all the surface layers containing silver and copper (or an excess of copper if copper is present in the lustre-free glaze) plus the thickness of the extreme surface metal-free layer when it exists; such parameter may bring information on the temperature and time of the final firing.
- The "total copper amount" and "total silver amount" sum up the quantity (expressed in $at.cm^{-2}$) of each of those two metallic elements contained in the lustre layers; this may be representative of the composition of the mixture applied by the potter to obtain the lustre.
- The "total copper volume fraction" and "total silver volume fraction" (in %) express the average volume fractions of copper and silver over the total thickness of the lustre; these values are the result of a combination of the two preceding parameters.
- The "main layer copper over silver volume fraction ratio" is the ratio between copper and silver volume fractions in the "main layer"; this parameter may give a representation of the potter intention to give a specific colouration to the lustre.

It is clear that, being the result of artisan handwork, the productions are not uniform and give rise to an important dispersion of the criteria within one given group. Nevertheless, some trends can be detected through a statistical study. The following figures **8-11** show a statistical evaluation of the preceding five criteria, comparing the different productions between them. The quantitative values used for the statistical computation are gathered in the extra figures **A1-A5** given in appendix 2. For each production group and each criterion, figures **8-11** give "box and whiskers" plots, showing for each group the median value (the value for which the numbers of objects with higher and lower values are equal), a box which delimitates the values adopted by 25 % and 75 % of the population, and the two maximum and minimum values ("whiskers") observed in the population. The number of objects (or different decorations when more than one colour is present on the same object) considered fro each production is indicated on the abscissa axe in each figure.

The groups are defined in a chrono-geographical manner, following the groups described in **table 1**. The Abbasid group is divided into 2 subgroups, depending on the nature of the glaze; "Abbasid-alk" concerns the fragments with an alkaline glaze and "Abbasid-Pb" the fragments with a lead glaze. The group "Abbasid ?" concerns the four specimens found in Fustat but with an uncertain attribution (third line of **table 1**) [**15**]. The groups "Fatimid ?", "Ayyubid ?" and "Nasrid ?" are for the specimens with an uncertain origin (also reported in **table 1**). The Hispano-Moresque production is divided into four sub-groups: "Cluny" for the plates of the Cluny Musée du Moyen-Age, "MNC" for the fragments of the Musée National de la Céramique, "Gubbio" for the fragments found in Gubbio (Italy) but clearly belonging to a Spanish Hispano-Moresque fabrication [**25**] and "Seville" for the fragments found in Seville and described in [**11**]. The Algerian group (La Qala de Banu Hammad) of **table 1** is not included in the analysis because the experimental results are not of sufficient reliability.

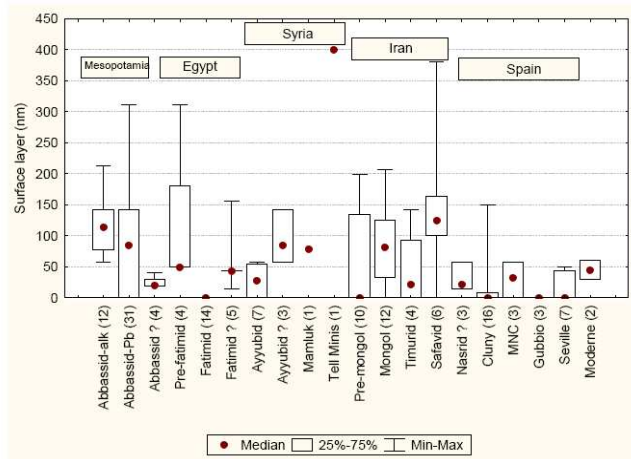

*Figure 8: statistical comparison of the productions from the viewpoint of the presence and thickness (in nm) of an extreme surface particle-free layer. See text for definition of the production groups.*

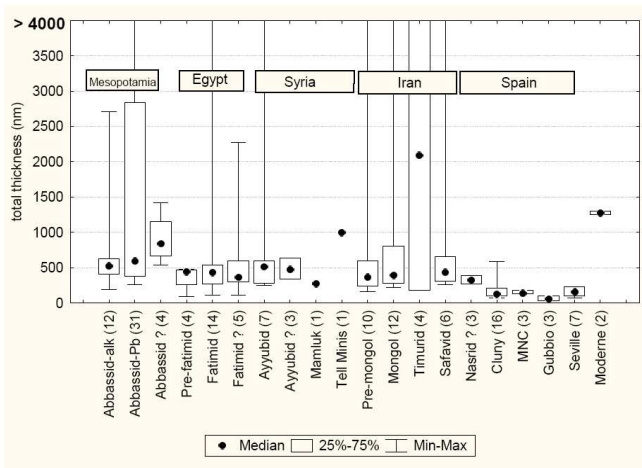

*Figure 9: statistical comparison of the productions from the viewpoint of the total thickness of the lustre, i.e. the sum of all layers containing Ag and/or Cu particles, plus the extreme surface metal-free layer when it exists. See text for definition of the production groups.*

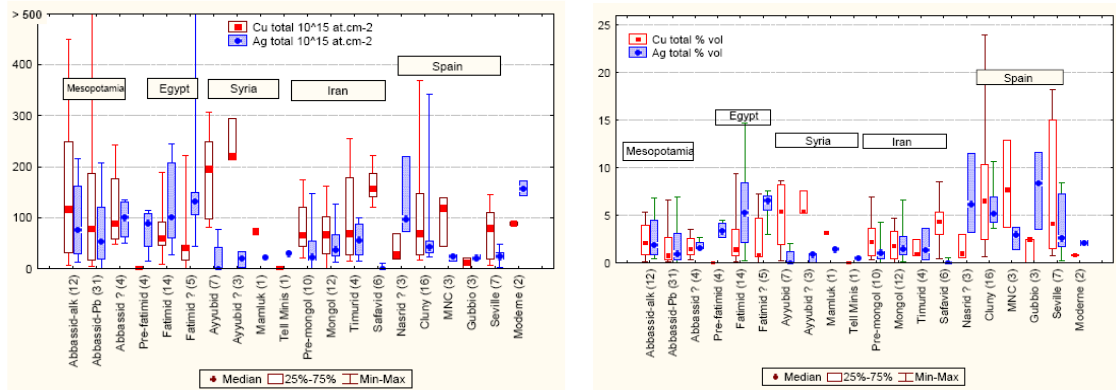

*Figure 10: comparison of the production from the viewpoint of the total Cu and/or Ag content of the lustre layers (left) and of the Cu and/or Ag volume fraction contained in the lustre layers. See text for definition of the production groups and of the volume contents.*

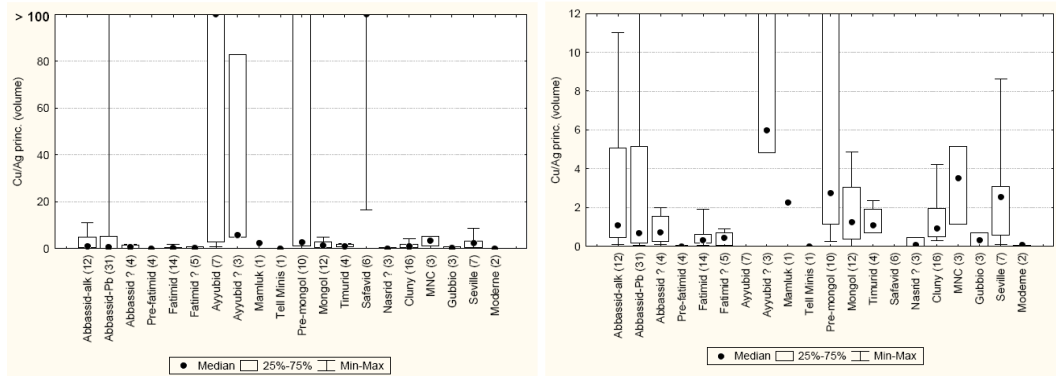

*Figure 11:* - *left: comparison of the productions from the viewpoint of the Cu/Ag volume fraction ratio in the main lustre layer (layer with the maximum Cu or Ag content) > 100 values are in fact generally infinite (Ag content is zero);*
*- right: a zoom on the low values of the left diagram.*
*See text for definition of the production groups and of the volume contents.*

If one focuses on those criteria, the following trends may be driven:

- The most ancient productions, namely corresponding to the Mesopotamian Abbasid sites and to the fragments found in Fustat (to be probably attributed to intermediate productions) exhibit a quite thick metal-free surface layer. This is no more true for the later productions (Fatimid), for which that surface layer is absent; its appears again in the Syrian and Iranian productions; in Spain, it becomes much thinner and even disappears for a large proportion of the Hispano-Moresque specimens; One may remind that the metal-free surface layer is always present on Italian majolica [**4**].
- For the Mesopotamian or Mesopotamian-like productions, the sum of the principal layer and gradient layers extends over a thickness reaching several micrometers, and this is also true for the modern artisan recreation (whose recipe was probably copied from the most ancient tradition). That sum is much smaller for the other productions, where the concerned thickness rarely exceeds 500 nm (except for Timurid shards) and is more often of the order of a few hundreds nanometres; it falls under 200 nm for the Spanish Hispano-Moresque objects [**26**]. One may remind that for Italian majolica that total thickness is even smaller [**4**].
- As for the copper and silver contents present (as metallic nanoparticles) in the lustre layers, they evidently depend on the final excepted aspect. The individual copper and silver contents in the lustre layers never exceed 14 at. %, except for the late (18$^{th}$ century) Hispano-Moresque objects where they may reach 40 at. %. From a general viewpoint, it may be said that those contents tend to increase with time, as the layers are becoming thinner. A comparison of the two diagrams of **figure 10** is interesting: the total amount of silver and/or copper is of the same order in nearly all the productions, but, owing to the differences in thickness, the final volume fraction of both metals is the lowest for the Mesopotamian lustres and the highest for the Hispano-Moresque lustres. Concerning the latest Hispano-Moresque productions it is worth noticing that, despite their very marked copper-red aspect, the lustre layers still contain noticeable quantities of silver, in opposition to a commonly admitted idea [**1**] which pretends that the potters of that period ceased to add silver in the paste mixture used for lustre elaboration.
- When one considers the Cu/Ag ratios, the diagrams of **figure 11** show that the Ag content is very small for a few number of productions (Ayyubid, pre-Mongol and Safavid) it is even zero for all the Safavid production, known to be constituted of copper-like coloured lustres on alkaline glaze. The zoom shown on the right part of **figure 11** gives an idea of the distribution of the Cu/Ag ratio around the value of unity, sometimes considered as a criterion to separate red-like lustres from golden-like lustres. It is now admitted [**14,15**] that the lustre colour is not entirely governed by the Cu/Ag ratio.
- A last point is worth to be underlined: the latest Spanish productions (18$^{th}$ century) are lutres with a strong copper-like shine, and the surface layers contain indeed high amounts of copper (20 to 25 % in volume fraction), although the surface layer still contain noticeable contents of silver [**26**]. The Iran Safavid lustres, produced at the same period show a comparable copper-like aspect and the surface layers contain also high amounts of copper but generally no silver. Both decorations are applied over transparent (tin-free) glazes (table 3), and the question might be raised of a possible influence of the late Spanish potter know-how on the renewal of the lustre fabrication in Iran during the 17$^{th}$ and the 18$^{th}$ century [**27**].

**5. Conclusions**

A study of metallic lustre decoration cannot be achieved without a comprehensive knowledge of their background, that is of the glazes and indirectly of the base terracotta.

First of all, if one focuses on the Abbasid production which can be considered as the reference because it is the first known metallic lustre on glaze, it constitutes a homogeneous entirety of terracotta. These are covered with glazes which can be either purely alkaline or alkaline with small lead amounts (PbO ≤ 6 wt %). The latter seem to correspond to a transition period of progressive introduction of lead and tin. For later periods, different ceramic technologies are used for the different studied productions. No correlation has been observed between the kind of ceramic body (baked clay or siliceous) and the glaze type (alkaline or lead-bearing alkaline). Yet, inside the same production site, a certain coherency can be shown in the results, specific of the production. Fatimid objects, in baked clay or siliceous paste, are systematically covered with lead-bearing alkaline glazes. The Ayyubid period shows a constant use of alkaline glazes affixed on siliceous paste. Hispano-Moresque objects are systematically elaborated from roasted clay and covered with lead glazes opacified with tin, to the noticeable exception of the end of the production (18$^{th}$ century) where the use of tin disappears.

About the metallic lustre, the following conclusions may be drawn:

For the Abbasid production, our reference, the use of copper and silver in various proportions leads to obtaining different colours or tones for the decoration. In other words the lustring mixture is a function of the desired result (speaking only on the Cu and Ag proportions, independently of the unknown possible influence of other compounds of the mixture). Yet, we suppose that the cooking process (temperature, reducing atmosphere) is relatively similar and mastered: systematic occurrence of a surface layer without copper and silver, similar thicknesses of the main layers and quasi-systematic occurrence of in-depth gradient. Each studied further production has its own technological specificity, shown here by the different layers and gradients of the decoration, which differentiates it from the Abbasid production: disappearance of the surface metal-free layer, variations in the metallic lustre layer thickness, variations in the copper and silver contents, etc. It seems thus that the metallic lustre technique has been adapted to a known local ceramic production, or, more precisely, to the nature of the glaze used locally (alkaline or lead-baring). It should be mentioned additionally that the latter plays also an important role on he firing process (melting or softening temperature of the glaze) and that it has an evident influence on the decoration microstructure.

We thus observe a constant back-and-forth between the esthetical desire and the technological choice, illustrated by the choice of copper and silver proportions. Variable proportions of Cu and Ag, already applied under Abbasids, are again found under Fatimids in Orient and at the beginning of the Hispano-Moresque period in Occident. The consequence is a chromatic variety of the decoration. For instance, the Fatimid lustre on coloured glazes use a major proportion of silver, allowing thus a good readability of the decoration.

On the contrary, a certain standardisation of the technology seems to be put in place for the Ayyubid period and also for the end of the Hispano-Moresque production, and this leads to some loss in the variety in the decoration colours. It is probably not a consequence of a losing of technology (the metal quantities in lustre are still as important or even larger) but rather an esthetical desire. As an example, one may consider the decorations produced at the end of the Hispano-Moresque period, which show a strong coppered aspect: the use of tin in the glaze is suddenly avoided. Without forgetting an always possible economical reason, one may consider an esthetical choice. The probably fashionable coppered aspect sought at that period agrees better with a cream-like background (colour of the body across the transparent glaze) than with a bright white background (colour of an opacified glaze).

**Acknowledgments**
This article is dedicated to the memory of Joseph Salomon, head of the AGLAE accelerator for the past 20 years. This study would not have existed without his active contribution.
Alex Kaczmarczyk[†] lent some of his fragments; Miguel Ruiz Jimenez gave spare modern fragments produced by himself [12].
That study was possible thanks to a post-doctorate fellowship supplied by C.N.R.S. to D. Chabanne. We are also indebted to the AGLAE team and the photographers of C2RMF for their technical support, and to Philippe Sciau (CEMES CNRS) who kindly supplied TEM photographs.

**References:**

[1] A. Caiger-Smith, *Lustre Pottery: Technique, Tradition and Innovation in Islam and the Western World* (Faber and Faber, London, U.K. 1985).
[2] R.B. Mason, *Shine like the sun: Lustre-Painted and Associated Pottery from the Medieval Middle east*, (Mazda Press, Costa Mesa, California and the Royal Ontario Museum, Toronto 2004), http://www.rbmason.ca/Shine/index.html.
[3] W.D.Kingery, P.Vandiver, *An Islamic Lusterware from Kashan, Ceramic masterpieces: Arts, Structure and technology* (Free Press, New-York 1986), pp. 111-121.
[4] G. Padeletti, G.M. Ingo, A. Bouquillon, S. Pages-Camagna, M. Aucouturier, S. Roehrs, P. Fermo, First-time observation of Mastro Giorgio masterpieces by means of non-destructive techniques, *Appl. Physics A* **83**, 475–483 (2006);
[5] D. Chabanne, *Le décor de lustre métallique des céramiques glaçurées ($IX^{eme}$-$XVII^{eme}$ siècles) : matériaux, couleurs et techniques. Principales étapes de diffusion d'une invention mésopotamienne*, PhD Thesis (Bordeaux 3 University, 3 juillet 2005).
[6] C.C. Piccolpasso, *The three books of the potter's art* (Victoria and Albert museum 1558).
[7] J. Perez-Arantegui, J. Molera, J. Larrea, A. Pradell, T. Vendrell-Saz, I. Borgia, B.G. Brunetti, F Cariati, P. Fermo, M. Mellini, A. Sgamelotti, C. Viti, Lustre pottery from the thirteenth to the sixteenth century: a nanostructured thin metallic film. *J. American Ceramics Society* **84**-2, 442-446 (2001),.
[8] S. Berthier, G. Padeletti, P. Fermo, A. Bouquillon, M. Aucouturier, E. Charron, V. Reillon, Lusters of renaissance pottery: Experimental and theoretical optical properties using inhomogeneous theories, *Appl. Phys. A*, **83**, 573–579 (2006).
[9] V. Reillon, S. Berthier, C. Andraud, New perspectives for the understanding of the optical properties of middle-age nano-cermets: The lustres, *Physica B*, **394**, 242-247 (2007);
V. Reillon, S. Berthier, C. Andraud, Optical properties of lustred ceramics: complete modelling of the actual structure, *Appl Phys A,* **100**, 901–910(2010)
[10] O. Bobin, M. Schvoerer, J.L. Miane, J.F. Fabre, Coloured metallic shine associated to lustre decoration of glazed ceramics: a theoretical analysis of the optical properties, *J. Non-Crystalline Solids* **332**, 28–34 (2003).

[11] A. Polvorinos del Rio, J. Castaing, M. Aucouturier, Metallic nano-particles in lustre glazed ceramics from the 15th century in Seville studied by PIXE and RBS, *Nuclear Instruments and Methods in Physics Research B* **249**, 596–600 (2006)

[12] http://www.miguelruizjimenez.com/

[13] M. Aucouturier, E. Darque-Ceretti, The surface of cultural heritage artefacts: physicochemical investigations for their knowledge and their conservation, *Chem. Soc. Rev.,* **36**, 1605–1621 (2007).

[14] D Hélary, E Darque-Ceretti, M. Aucouturier, Contemporary Golden-Like Lusters on Ceramics: Morphological, Chemical, and Structural Properties, *J. American Ceramic Society*, **88**, 3218–3221 (2005).

[15] E. Darque-Ceretti, D. Hélary, A. Bouquillon, M. Aucouturier, Gold like lustre: nanometric surface treatment for decoration of glazed ceramics in ancient Islam, Moresque Spain and Renaissance Italy, *Surf. Engineering*, **21**, 352-358 (2005).

[16] A. Guinier, *Théorie et technique de la radiocristallographie,* (Dunod, Paris 1964).

[17] T. Calligaro, J.-C. Dran, J. Salomon, Ion beam microanalysis, in: *Non-destructive microanalysis of cultural heritage materials*, K. Janssens and R. Van Grieken ed. , Comprehensive Analytical Chemistry XLII, (Elsevier, Amsterdam 2005), pp. 227-276.

[18] J.A. Maxwell, J.L. Campbell, W.J. Teesdale, The Guelph PIXE software package, *Nuclear Instruments and methods in physics research*, *B* **43**, 218-230 (1988).

[19] M. Mayer, *SIMNRA*, © Max-Planck-Institut für Metallphysik, 1997–1998, http://www.rzg.mpg.de/

[20] D. Chabanne, O. Bobin, M. Schvoerer, C. Ney, Metallic lustre of glazed ceramics: evolution of decorations in search for discriminating elements, *34$^{th}$ internat. Symposium on Archaeometry,* Zaragoza May 2004, pub. on-line: http://www.dpz.es/ifc/libros/libros.htm.

[21] S. Lahlil, A. Bouquillon, G. Morin, L. Galoizy, C. Lorre, Relationship between the colouration and the firing technology used to produce Susa glazed ceramics of the end of the Neolithic period, *Archaeometry*, **51**/5, 774-790 (2009).

[22] D. Arnold, *Studien zur altägyptischen Keramik* (Verlag Philipp von Staben (SDAIK 9), Mainz am Rhein, 1981).

[23] F. Hamer, J. Hamer, *The potter's dictionary of materials and techniques*, (A et C Black, London, 1997) 406 p.

[24] S. Padovani, D. Puzzovio, C. Sada, P. Mazzoldi, I. Borgia, A. Sgamellotti, B.G. Brunetti, L. Cartechini, F. D'acapito, C. Maurizio, F. Shokoui, P. Oliaiy, J. Rahighi, M. Lamehi-rachti, E. Pantos, XAFS study of copper and silver nanoparticles in glazes of medieval middle-east lusterware (10th–13th century), *Appl. Phys. A* **83**, 521–528 (2006)

[25] D. Chabanne, A. Bouquillon, M. Aucouturier, X. Dectot, G. Padeletti, Physico-chemical analyses of Hispano-Moresque lustred ceramic: a precursor for Italian majolica?, *Appl. Phys. A,* **92**, 11–18 (2008)

[26] X. Dectot, Les *céramiques hispaniques (XIIe-XVIIIe siècle)*, Musée national du Moyen Âge, Thermes et Hôtel deCluny (Paris, Réunion des Musées nationaux, 2007), p. 160

[27] S. Makariou, Et retour vers l'Orient ? In : *Reflets d'or. D'Orient en Occident la céramique lustrée*, exhibition catalogue, X. Dectot, S. Makariou and D. Miroudot exhibition curators (Paris, Réunion des Musées nationaux, April 2008), p.107.

*Table 2: average composition of the terra cotta in wt %, measured by PIXE.*
*Siliceous pastes are on grey background. S.D. = standard deviation*

| Provenance | | Na$_2$O | MgO | Al$_2$O$_3$ | SiO$_2$ | P$_2$O$_5$ | SO$_3$ | Cl | K$_2$O | CaO | TiO$_2$ | MnO | Fe$_2$O$_3$ |
|---|---|---|---|---|---|---|---|---|---|---|---|---|---|
| Abbassid | mean | 2.26 | 7.08 | 12.56 | 45.91 | 0.20 | 1.90 | 0.78 | 1.37 | 19.64 | 0.59 | 0.14 | 7.35 |
| | *S.D.* | *1.70* | *0.73* | *0.91* | *2.83* | *0.08* | *2.80* | *1.59* | *0.38* | *1.24* | *0.21* | *0.02* | *0.51* |
| Pré-Fatimid | mean | 2.16 | 3.50 | 11.62 | 42.95 | 0.47 | 3.78 | 2.02 | 1.38 | 22.57 | 1.06 | 0.10 | 7.83 |
| | *S.D.* | *1.5* | *0.9* | *2.3* | *7.6* | *0.4* | *4.2* | *2.3* | *0.3* | *6.6* | *0.3* | *0.0* | *1.3* |
| Fatimid | mean | 1.39 | 3.56 | 12.34 | 47.60 | 0.50 | 2.09 | 0.66 | 1.20 | 21.81 | 0.98 | 0.09 | 7.18 |
| | *S.D.* | *0.34* | *0.27* | *0.76* | *3.68* | *0.20* | *1.47* | *0.55* | *0.26* | *3.13* | *0.22* | *0.01* | *0.75* |
| Fatimid | mean | 4.81 | 1.18 | 7.79 | 75.50 | 0.25 | 1.50 | 0.93 | 1.08 | 3.66 | 0.42 | 0.04 | 1.60 |
| | *S.D.* | *1.11* | *0.79* | *1.02* | *3.15* | *0.13* | *1.52* | *0.52* | *0.23* | *1.31* | *0.06* | *0.05* | *0.20* |
| Ayyubid | mean | 3.59 | 2.61 | 3.32 | 78.48 | 0.51 | 1.86 | 0.44 | 1.59 | 5.54 | 0.21 | 0.03 | 1.55 |
| | *S.D.* | *0.57* | *0.59* | *1.36* | *5.03* | *0.69* | *2.30* | *0.40* | *0.93* | *2.02* | *0.19* | *0.01* | *0.30* |
| Pré-Mongol | mean | 2.83 | 1.06 | 8.13 | 77.48 | 0.25 | 2.43 | 0.25 | 1.84 | 2.86 | 1.04 | 0.02 | 1.40 |
| | *S.D.* | *0.33* | *0.34* | *2.36* | *3.46* | *0.17* | *1.91* | *0.10* | *0.37* | *1.07* | *0.36* | *0.01* | *0.47* |
| Mongol | mean | 2.97 | 1.58 | 7.72 | 77.91 | 0.36 | 1.27 | 0.19 | 1.59 | 3.82 | 0.89 | 0.03 | 1.36 |
| | *S.D.* | *0.30* | *0.48* | *1.96* | *3.64* | *0.28* | *1.68* | *0.11* | *0.31* | *1.23* | *0.26* | *0.01* | *0.30* |
| Timurid | mean | 1.01 | 3.29 | 12.60 | 50.48 | 0.12 | 0.98 | 0.20 | 2.63 | 21.58 | 0.70 | 0.11 | 5.98 |
| | *S.D.* | *0.02* | *0.05* | *0.86* | *0.10* | *0.17* | *0.34* | *0.22* | *0.03* | *0.26* | *0.06* | *0.00* | *0.41* |
| Algeria | mean | 0.91 | 2.32 | 14.06 | 40.61 | 0.38 | 1.18 | 0.17 | 0.61 | 30.43 | 1.07 | 0.09 | 7.30 |
| | *S.D.* | *0.11* | *0.40* | *2.95* | *3.62* | *0.41* | *0.41* | *0.08* | *0.16* | *7.33* | *0.36* | *0.02* | *0.74* |
| Hispano-Moresque | mean | 0.80 | 3.05 | 14.16 | 44.49 | 0.15 | 3.23 | 0.24 | 2.79 | 23.49 | 0.77 | 0.10 | 5.93 |
| | *S.D.* | *0.18* | *0.47* | *0.94* | *3.08* | *0.26* | *1.75* | *0.13* | *0.89* | *3.40* | *0.30* | *0.07* | *0.42* |

*Table 3: average composition of the glazes in wt %. measured by PIXE.*
*discriminating elements for opaque lead glazes are **in bold**. Colouring additions (Co, Cu) are not mentioned.*

| Provenance | Type | Na$_2$O | MgO | Al$_2$O$_3$ | SiO$_2$ | P$_2$O$_5$ | SO$_3$ | Cl | K$_2$O | CaO | TiO$_2$ | MnO | Fe$_2$O$_3$ | SnO$_2$ | PbO |
|---|---|---|---|---|---|---|---|---|---|---|---|---|---|---|---|
| Abbassid 1* | transparent | 3.10 | 2.30 | 2.99 | 72.59 | 0.15 | 4.26 | 0.48 | 4.86 | 6.42 | 0.16 | 0.56 | 1.10 | 0.15 | 0.25 |
| Abbassid 2* min | | 3.85 | 3.41 | 2.29 | 72.17 | 0.20 | 0.85 | 0.53 | 4.70 | 5.55 | 0.12 | 0.27 | 0.90 | **3.20** | 1.70 |
| Abbassid 2* max | opaque | 0.63 | 1.67 | 2.82 | 51.55 | 1.02 | 8.48 | 0.65 | 2.28 | 6.21 | 0.06 | 0.12 | 0.72 | **12.78** | 8.00 |
| Pre-Fatimid white | opaque | 1.41 | 0.08 | 2.29 | 43.81 | 0.00 | 0.00 | 0.21 | 3.12 | 0.65 | 0.26 | 0.01 | 0.51 | **18.46** | **29.00** |
| Fatimid (2 spec.) | opaque | 1.70 | 0.10 | 1.90 | 48.45 | 0.02 | 0.00 | 0.14 | 3.27 | 0.85 | 0.33 | 0.01 | 0.54 | **7.85** | **34.60** |
| Fatimid (2 spec.) | opaque | 0.86 | 0.09 | 1.56 | 38.54 | 0.00 | 0.00 | 0.29 | 1.55 | 0.57 | 0.34 | 0.01 | 0.44 | **14.54** | **40.75** |
| Ayyubid | transparent | 8.54 | 2.54 | 1.89 | 74.68 | 0.43 | 1.06 | 0.44 | 2.61 | 5.88 | 0.15 | 0.08 | 1.40 | 0.00 | 0.01 |
| Mamluk | transparent | 5.03 | 3.56 | 1.44 | 73.15 | 0.13 | 0.36 | 0.12 | 4.57 | 6.99 | 0.09 | 0.03 | 2.18 | 0.01 | 0.22 |
| Pre-Mongol | opaque | 4.29 | 1.89 | 1.82 | 55.06 | 0.04 | 0.36 | 0.27 | 2.04 | 3.14 | 0.11 | 0.03 | 0.73 | **9.32** | **20.21** |
| Mongol | opaque | 3.95 | 1.65 | 2.24 | 59.23 | 0.00 | 0.62 | 0.27 | 3.10 | 4.07 | 0.18 | 0.03 | 1.00 | **6.77** | **16.39** |
| Timurid | opaque | 2.16 | 1.41 | 2.22 | 59.31 | nd | nd | 0.14 | 3.42 | 2.94 | 0.09 | 0.02 | 0.54 | **8.15** | **18.37** |
| Safavid | transparent | 7.25 | 3.10 | 2.24 | 78.03 | 0.07 | 0.23 | 0.11 | 1.75 | 5.56 | 0.12 | 0.06 | 1.11 | 0.00 | 0.01 |
| Pre-Nasrid and Nasrid | opaque | 1.20 | 0.49 | 0.93 | 45.15 | 0.06 | 0.00 | 0.05 | 2.32 | 1.42 | 0.07 | 0.01 | 0.40 | **6.70** | **40.96** |
| Hispano-Moresque 1 (<17[th] cent.) | opaque | 0.73 | 0.30 | 1.85 | 46.37 | nd | 0.07 | 0.17 | 5.42 | 2.49 | 0.06 | 0.01 | 0.22 | **5.38** | **36.43** |
| Hispano-Moresque 2 (18[th] cent.) | transparent | 0.57 | 0.11 | 1.31 | 55.73 | nd | 0.08 | 0.19 | 5.02 | 1.35 | 0.06 | 0.01 | 0.40 | 0.07 | 33.47 |

\* Abbassid glazes are divided into two categories: "Abbassid 1" have an homogeneous composition; "Abbassid 2" are very scattered and minimum and maximum measured values are indicated

Figure captions

Figure 1: fabrication principle and structure of a glazed ceramic with lustre decoration

Figure 2: examples of the studied objects (whole pieces or fragments): Abbasid period (A & B, © C2RMF, D. Bagault), Fatimid period (C & D, © C2RMF, D. Bagault) and Hispano-Moresque period (E et F, © Musée du Moyen-Âge, Cluny)

Figure 3: origin and dating of the studied objects

Figure 4: observation of a lustred ceramic at different scales: (a)optical microscope; (b) conventional SEM; (c) HR-SEM; (d) AFM; (e) TEM

Figure 5: grazing X-ray diffraction on a modern lustre surface ($\alpha$ = incidence angle)

Figure 6: experimental (curves with noise) and simulated RBS spectra on a specimen of lustred ceramic; (a) simulation of the glaze covered with a layer containing Cu and Ag; (b) addition of a layer containing neither copper nor silver; (c) insertion of an intermediary layer with a Cu and/or Ag gradient; (d) simplified result of the simulation; (e) TEM image of a cross section of a similar specimen (© CEMES-CRPAA, P. Sciau).

Figure 7: three lustre structures observed in TEM and corresponding results by RBS analysis simulation. TEM micrographs kindly provided by P. Sciau, CEMES-CNRS, Toulouse. Concentrations are in at. %.

Figure 8: statistical comparison of the productions from the viewpoint of the presence and thickness (in nm) of a particle-free layer at the extreme surface. See text for definition of the production groups.

Figure 9: statistical comparison of the productions from the viewpoint of the total thickness of the lustre, i.e. the sum of all layers containing Ag and/or Cu nanoparticles, plus the extreme surface metal-free layer when it exists. See text for definition of the production groups.

Figure 10: comparison of the production from the viewpoint of the total Cu and/or Ag content of the lustre layers (left) and of the Cu and/or Ag volume fraction contained in the lustre layers. See text for definition of the production groups and of the volume contents.

Figure 11: - left: comparison of the productions from the viewpoint of the Cu/Ag volume content ration in the main lustre layer (layer with the maximum Cu or Ag content) > 100 values are in fact generally infinite (Ag content is zero);
- right: a zoom on the low values of the left diagram.
See text for definition of the production groups and of the volume contents.

# APPENDIX 1
## The studied ceramics

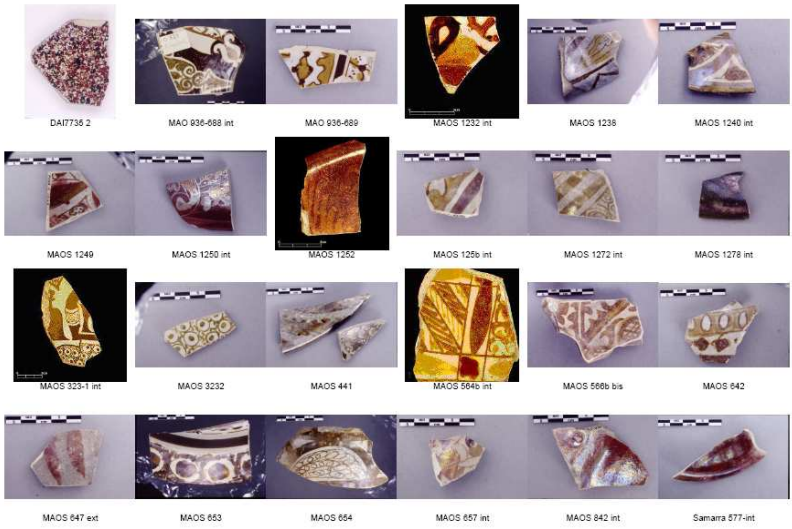

**Abbasids**

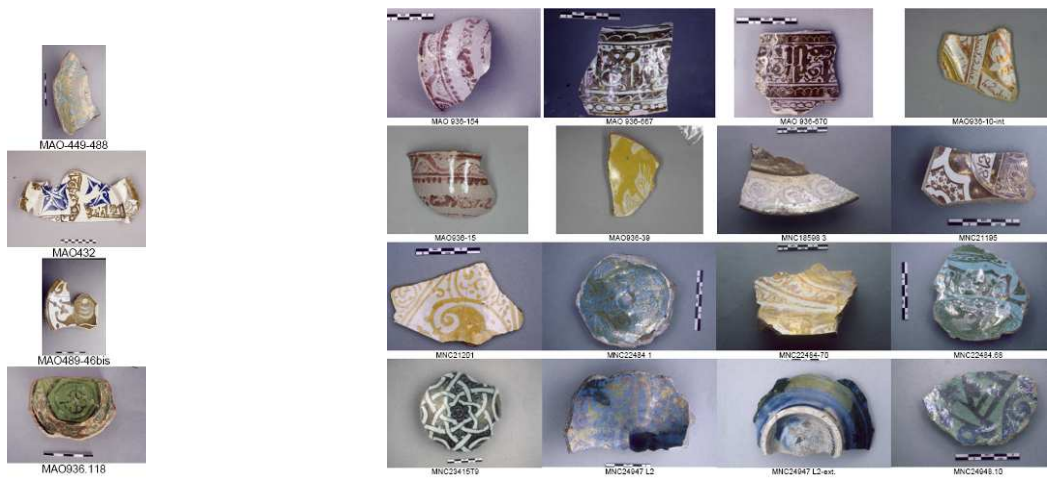

**EGYPT, pre-fatimids**  **EGYPT, Fatimids**

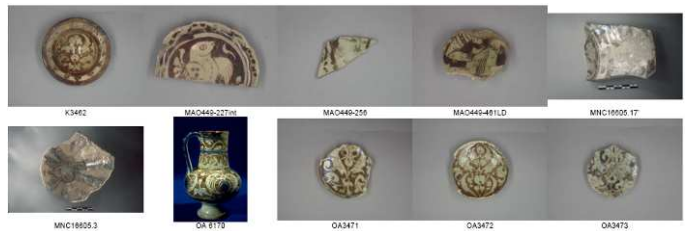
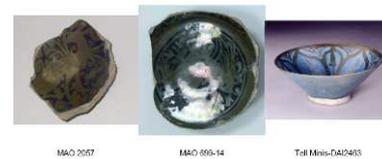

**SYRIA, Ayyubids**  **SYRIA, Mamluk and Tel Minis**

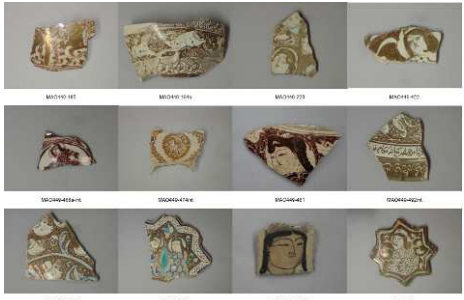
**IRAN, pre-mongols**

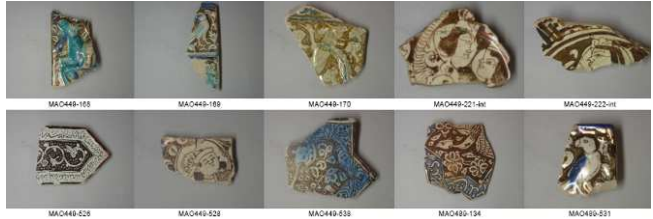
**IRAN, Mongols**

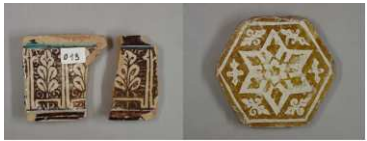
**IRAN, Timurids**

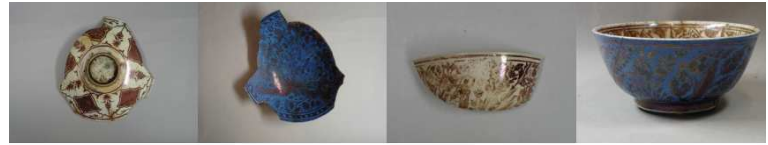
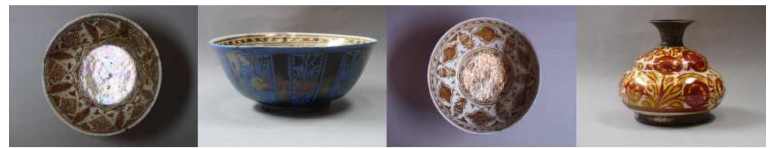
**IRAN, Safavids**

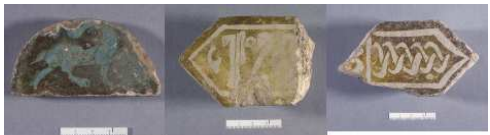
**ALGERIA**

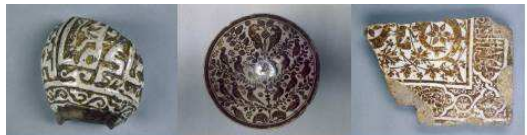
**ISLAMIC SPAIN, 12th – 14th cent.**

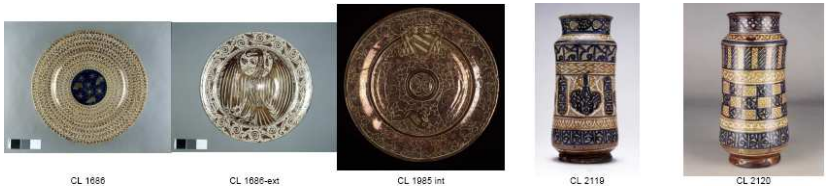
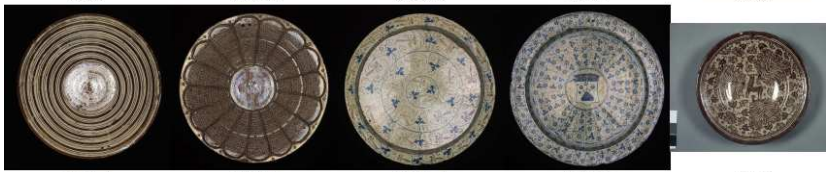
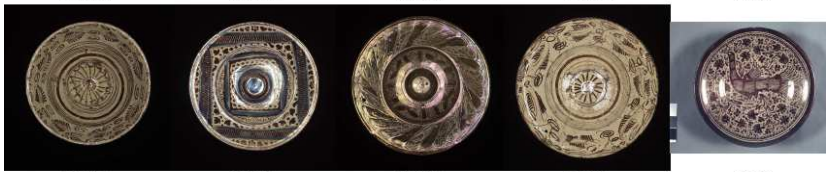
**SPAIN, Valencia region, Cluny museum**

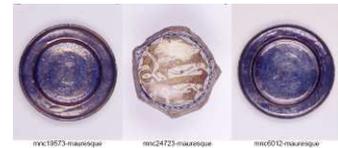
**SPAIN, Valencia region Musée National de la Céramique**

# APPENDIX 2
## Quantitative individual values used for the statistical figures 8 to 12
See text for definition of the production groups

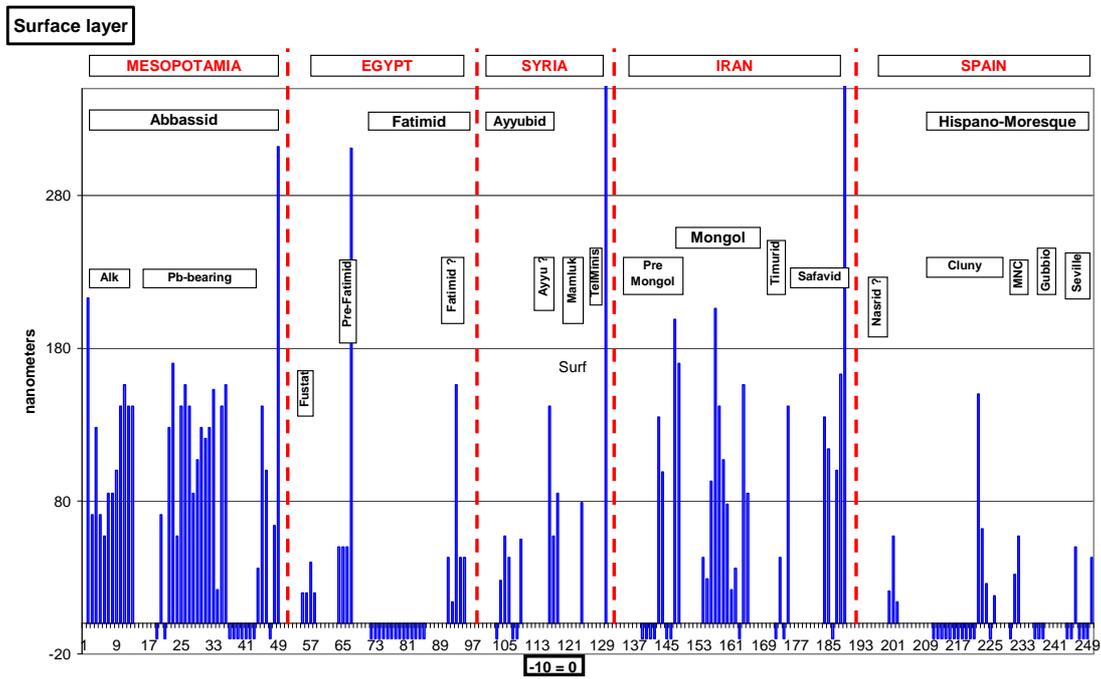

**Figure A1:** thickness of the silver and copper-free surface layer for each lustre.
("-10" means the absence of surface layer)

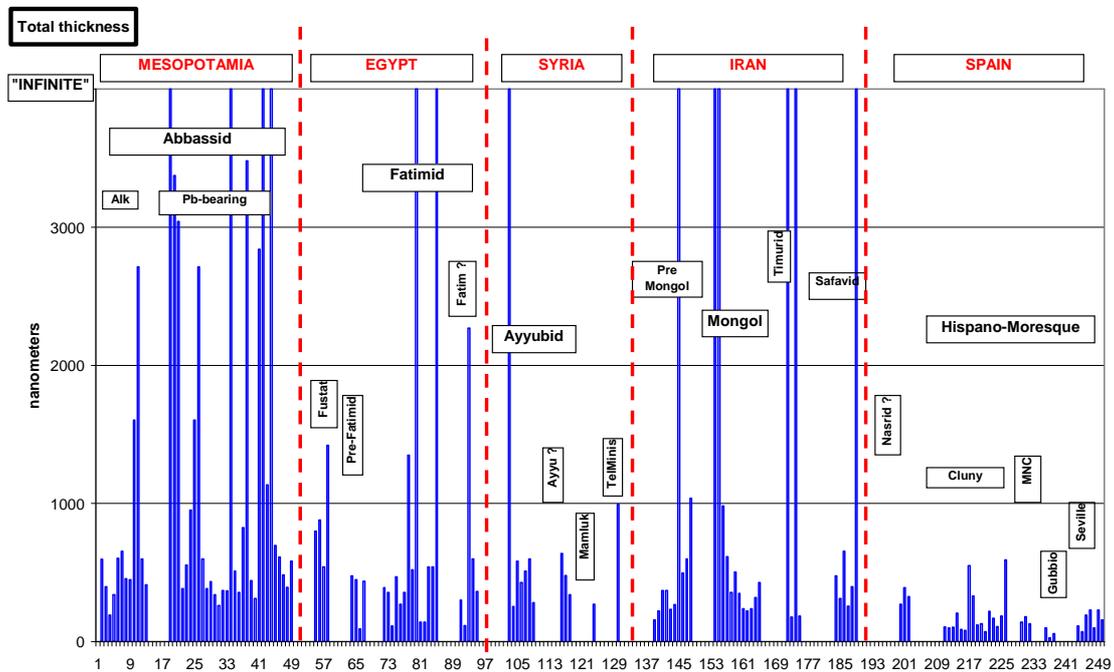

**Figure A2:** total thickness of each lustre (sum of the surface metal-free layer and of the layers containing metallic copper and silver nanoparticles).
"Infinite" means a thickness larger than the thickness explored by RBS (> 5 μm)

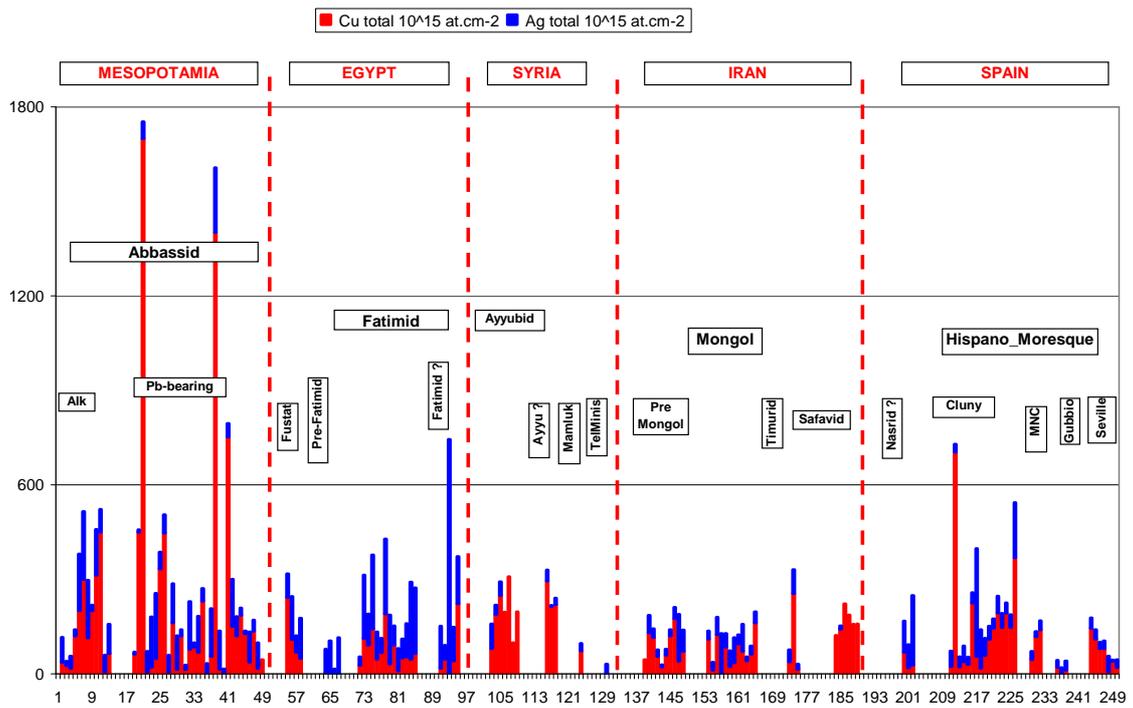

**Figure A3**: total copper and silver contents in each lustre, expressed in at.cm$^{-2}$

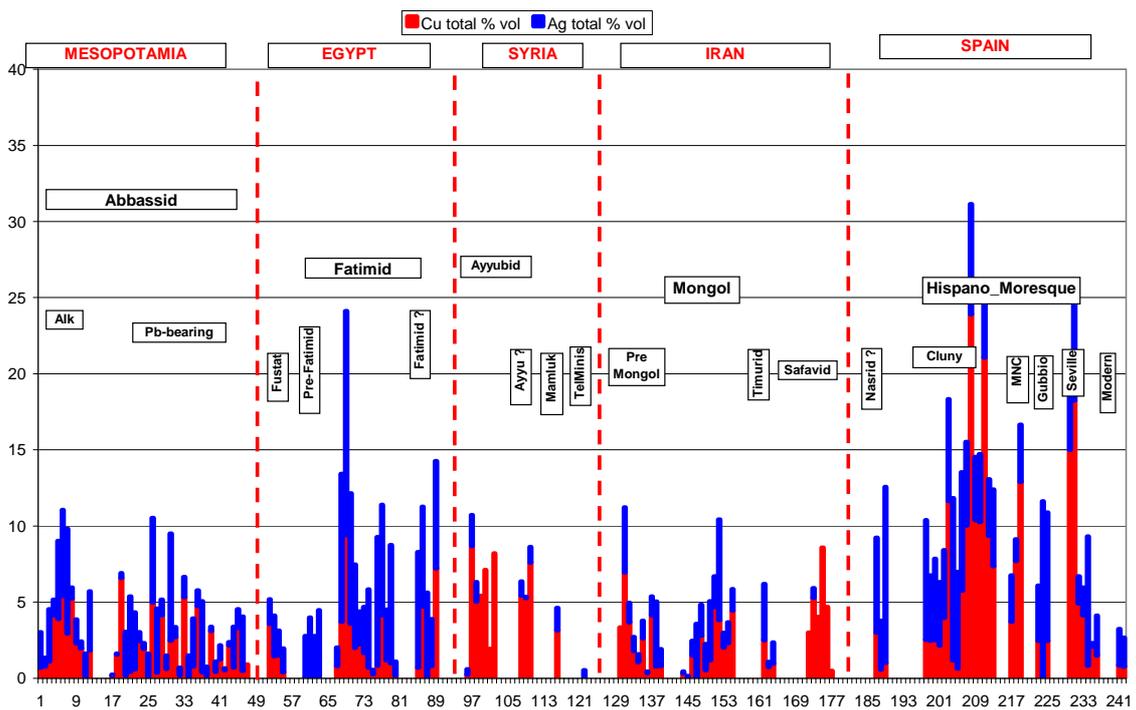

**A4:** copper and silver volume fraction (in %) in each lustre layer (see text for definition of the volume fraction)

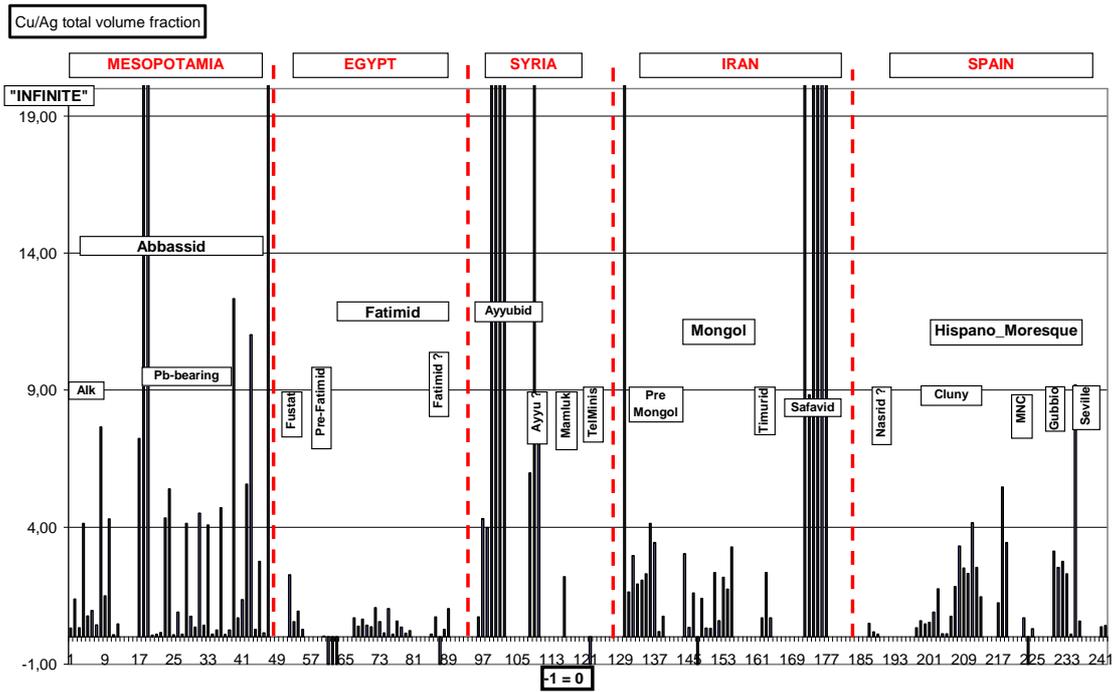

**A5:** Cu/Ag volume fraction ratio for each lustre.
"-1" means Cu/Ag= 0 (no copper); "Infinite" means no silver